\documentclass[aps,prl,twocolumn,groupedaddress]{revtex4}

\usepackage{graphicx}

\usepackage{amssymb}

\begin{document}

\title{Primary role of the barely occupied states in the charge density wave formation of NbSe$_{2}$}

\author{D. W. Shen$^{1*}$\email{dwshen@fudan.edu.cn}, Y. Zhang$^1$, L. X. Yang$^1$, J.
Wei$^1$, H. W. Ou$^1$, J. K. Dong$^1$, C. He$^1$, B. P. Xie$^1$, B.
Zhou$^1$, J. F. Zhao$^1$M. Arita$^2$, K. Shimada$^2$, H.
Namatame$^2$, M. Taniguchi$^2$, J. Shi$^3$, and D.L.
Feng$^1$\email{dlfeng@fudan.edu.cn}}

\email{dwshen@fudan.edu.cn;dlfeng@fudan.edu.cn}

\affiliation{$^1$Department of Physics, Surface Physics
Laboratory(National Key Laboratory) and Advanced Materials
Laboratory, Fudan University, Shanghai 200433, P. R. China}

\affiliation{$^2$Hiroshima Synchrotron Radiation Center and Graduate
School of Science, Hiroshima University, Hiroshima 739-8526, Japan}

\affiliation{$^3$School of Physics, Wuhan University, Wuhan, 430072,
P. R. China}

\date{\today}

\begin{abstract}

NbSe$_{2}$ is a prototypical charge-density-wave (CDW) material,
whose mechanism remains mysterious so far. With angle resolved
photoemission spectroscopy, we recovered the long-lost nesting
condition over a large broken-honeycomb region in the Brillouin
zone, which consists of six saddle band point regions with high
density of states (DOS), and large regions away from Fermi surfaces
with negligible DOS at the Fermi energy. We show that the major
contributions to the CDW formation come from these barely occupied
states rather than the saddle band points. Our findings not only
resolve a long standing puzzle, but also overthrow the conventional
wisdom that CDW is dominated by regions with high DOS.

\end{abstract}

\maketitle

2H structured niobium diselenide (NbSe$_{2}$) is one of the most
studied materials for its  prototypical superconductivity (SC) and
two dimensional charge density wave (CDW)
 \cite{webofscience,Revolinski,Harper1975,ThStraub,shinScience,Valla2004}.
Based on experiments conducted on  NbSe$_{2}$,  understanding on
some of the most basic properties of SC and CDW was reached, such as
the internal electronic structures of the vortex cores \cite{Choi},
the anisotropic s-wave \cite{Choi,shinScience,Valla2004} and
multi-band SC
 \cite{Sonier,Boaknin,Fletcher,Narzykin}, the anisotropic
electron-phonon interaction \cite{Valla2004,Kiss}, and the phonon
softening at the CDW transition \cite{MonctonPRB}. Nevertheless, the
very mechanism of the CDW in NbSe$_{2}$ itself has been mysterious
and controversial for over three decades
 \cite{RicePRL,ThStraub,Rossnagel,Johannes,Kiss}, even though it is
one of the very first two dimensional CDW materials discovered
 \cite{WilsonAP}.

In a typical picture of CDW, there are usually parallel Fermi
surface (FS) sections, which guarantee a large number of electrons
could be scattered from one parallel side to the other by softened
phonons with a fixed momentum transfer corresponding to the ordering
wavevector(s). As a result, the charge susceptibility at the
ordering wavevector(s) will be enhanced, and CDW instability could
be induced in collaboration with lattice. This gives the so called
nesting condition of CDW. Similarly, saddle band points, where the
singularities of density of states (DOS) are located, were proposed
to cause CDW as well \cite{RicePRL}. Recently, the authors discussed
another so-called Fermi-patch CDW mechanism for some polaronic
systems, where large DOS at $E_F$ hovers over an extended area of
the Brillouin zone (BZ) and work collectively to favor certain CDW
order \cite{ShenTMD}. However, the CDW in NbSe$_{2}$ cannot be
explained by any of these scenarios:  neither can its ordering
wavevectors match the FS sections
 \cite{ThStraub,Rossnagel,Johannes}, nor match the saddle band
points \cite{Rossnagel,Johannes}, even though some experiments
showed that the CDW gap does open near the saddle band points
 \cite{LiuPRL}; and there are no Fermi patches. The latest experiment
identified the CDW gap in the hashed circles of
Fig.\,\ref{Auto}(\textbf{a}), some of which fulfill the nesting
condition \cite{Kiss}. However, there are still some regions that do
not. The inapplicability of these common models for CDW in such a
classic compound suggests that our current understanding of CDW need
to be amended.

% summary of what we have done

All the existing CDW mechanisms are centered on the regions with
high DOS at $E_F$ in the BZ. In this Letter, we show that the CDW
related spectral weight suppression with large energy scale occurs
in a large broken-honeycomb region mainly away from the FS's,
including a large momentum area where the DOS is very small at
$E_F$. Recovery of the region makes the nesting conditions naturally
fulfilled, which explains the previous controversies. Particularly,
we prove that it is the states in these low-DOS regions that play a
primary role in the CDW formation. Such an unexpected finding
overthrows the conventional wisdom that CDW is dominated by the
regions with high DOS. The electronic states in the entire BZ have
to be examined no matter how low the DOS they possess.

\begin{figure}[htbp]
\includegraphics[width=8cm]{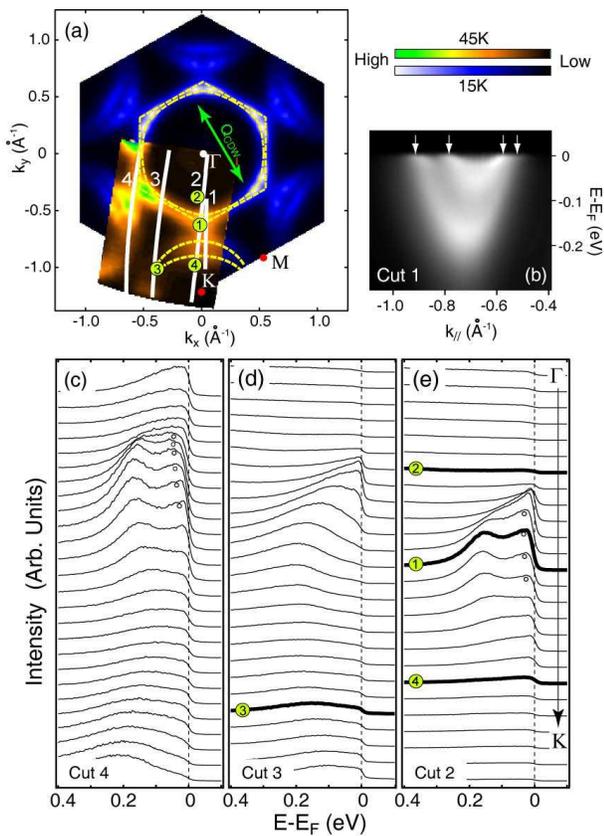}
\caption{(Color online) (\textbf{a}) Photoemission intensity maps
(integrated within 20meV around $E_F$) at the CDW state (15K) and
normal state (45K) are compared. The image taken at 15K is 6-fold
symmetrized. (\textbf{b}) The photoemission intensity along Cut 1
marked in panel (\textbf{a}). (\textbf{c})-(\textbf{e}) The typical
ARPES spectra for NbSe$_{2}$ taken at normal state (45K) along the
corresponding cuts. The thicker curves are the spectra at the
numbered momenta in panel (\textbf{a}), and the open circles
represent the location of the flat band. } \label{EDC}
\end{figure}

% experimental part
High quality NbSe$_{2}$ single crystals were synthesized by the
vapor-transport technique. The data were collected with 21.2eV
photons from Helium discharge lamp and BL9 of Hiroshima synchrotron
radiation center by the Scienta R4000 electron analyzers. The
angular resolution is $0.3^\circ$ and the total energy resolution is
$\sim10\,meV$. In the temperature dependence studies, measurements
were conducted in a cyclic way to guarantee no aging effects in the
spectra. All experiments were performed in the ultra-high vacuum
(better than 3$\times 10^{-11}\,mbar$ in the helium lamp system and
$\sim5\times 10^{-11}\,mbar$ at BL-9) and in a short time.

% Results: general picture
Photoemission intensity maps integrated within $\pm$10\,meV around
the Fermi energy ($E_F$) at the normal (45K) and CDW states (15K)
are compared in Fig.\,\ref{EDC}(\textbf{a}). Three FS pockets, one
hexagon around \textbf{$\Gamma$} and two around \textbf{K} point can
be directly distinguished. Furthermore, because of the coupling
between two NbSe$_2$ layers in a unit cell, two split bands can be
identified [see Fig.\,\ref{EDC}(\textbf{b})], which gives
double-walled FS's \cite{Mattheiss,Corcoran,Johannes}. Although
there are long straight sections in the \textbf{$\Gamma$} pockets,
none of them could be connected by the CDW wavevector
\textbf{Q$_{i}$}$\sim$\textbf{a$_{i}$}*/3=0.688
${\AA}^{-1}$(\emph{i=1,2,3}) \cite{Moncton1}.

%explain NbSe2 is different from NaTaS2, it has very well defined
%Fermi surface and no Fermi patch. Saddle large, flat band

Figs.\,\ref{EDC}(\textbf{c-e}) illustrate the normal state spectra
of NbSe$_{2}$ along three cuts across FS's.  None of them exhibits a
sharp quasiparticle-like lineshape, even at the Fermi crossings.
However, they do show clear dispersions, which make the FS's well
defined. These have been found to be typical signatures of polaronic
systems, such as La$_{1.2}$Sr$_{1.8}$Mn$_{2}$O$_{7}$ and
K$_{0.3}$MoO$_{3}$, where the weight of the quasiparticle peak is
vanishingly small, and its dispersion is renormalized to the
vicinity of the FS. However, the majority of states are distributed
to much higher energies as incoherent spectral weight by scattering
of multi-phonons\cite{LSMO,KMO}, which interestingly, still follows
the bare band dispersion. We note that in NbSe$_{2}$ the spectral
weight at $E_F$ quickly drops when away from the FS's, unlike a
sibling compound 2H-Na$_{x}$TaS$_{2}$, where large spectral weight
at $E_F$ is observed almost over the entire BZ \cite{ShenTMD}.
Moreover, midway between \textbf{$\Gamma$} and \textbf{K}, where the
saddle band points stand, we indeed observe the flat dispersion as
marked by the open circles in Figs.\,\ref{EDC}(\textbf{c}) and
\ref{EDC}(\textbf{e}). Even though the spectral centroid is below
$E_F$, the DOS is quite high here.

\begin{figure*}[htbp]
\includegraphics[width=13cm]{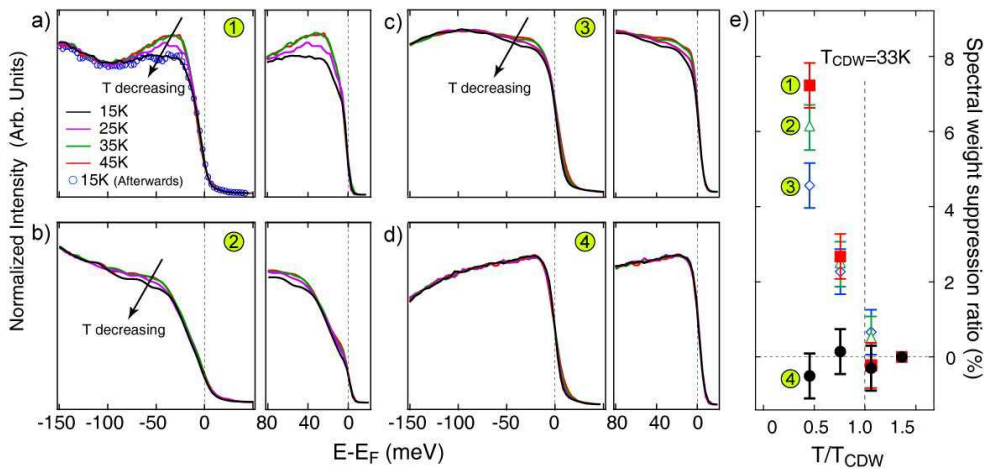}
\caption{(Color online) (\textbf{a})-(\textbf{d}) Typical
temperature dependence of the original (left) and temperature
broadening effects removed (right) spectra in the normal and CDW
states at momenta No.\textbf{1-4} marked in
Fig.\,\ref{EDC}(\textbf{a}) respectively (see text for details).
(\textbf{e}) The spectral weight suppression ratios (compared to
spectra taken at 45K and over ($E_F$-80\,meV, $E_F$)) for different
temperatures in fig (\textbf{a-d}) respectively.}

\label{Tdependence}% \vspace*{-0.5cm}
\end{figure*}

The intensity maps do not show any typical CDW effects, such as the
folding of the FS's. However, through careful examination of the
spectra, the suppression upon CDW is observed in various regions of
the BZ. Several representative spectra are plotted in
Figs.\,\ref{Tdependence}(\textbf{a-d}) for illustration. The
corresponding momenta are marked in Fig.\,\ref{EDC}(\textbf{a}) and
their corresponding spectra have been highlighted in
Figs.\,\ref{EDC}(\textbf{d-e}). The left panel of
Fig.\,\ref{Tdependence}(\textbf{a}) shows a spectral lineshape
evolution as a function of temperature near the saddle band point.
The spectra taken at 45K and 35K, which are both above the CDW
transition temperature 33K, nearly coincide with each. Once entering
the CDW state, the spectrum is suppressed noticeably over about
80\,meV below $E_F$ while the leading edge shifts negligibly. This
suppression energy scale is much higher than the $k$$_{B}$$T$$_{c}$
of 3\,meV, which is the energy scale of CDW effects on the
quasiparticles. Such low energy effects might be passed to the
incoherent spectral weight at high energies by multi-phonon
scattering processes as observed here.

The suppression at saddle point agrees with previous observations
\cite{Kiss}. However, the main discovery here is that such CDW
related suppression also occurs over large momentum regions inside
the \textbf{$\Gamma$} pocket [Fig.\,\ref{Tdependence}(\textbf{b})]
and around the M point[Fig.\,\ref{Tdependence}(\textbf{c})], though
the spectral weights at $E_F$ are low there [see
Figs.\,\ref{EDC}(\textbf{d-e})]. Whereas, for the spectra taken
inside the \textbf{K} pocket[Fig.\,\ref{Tdependence}(\textbf{d})],
no such suppression is observed. To show the suppression is not due
to thermal broadening, all spectra are divided by the resolution
convoluted Fermi functions at corresponding temperatures and then
multiplied by that of 10K, before they are compared in the right
panels of Figs.\,\ref{Tdependence}(\textbf{a-d}). While the
suppression at momenta No.\textbf{1-3} is observed upon CDW
formation, it is absent at momentum No.\textbf{4}. This is further
illustrated by the spectral weight suppression ratios in
Fig.\,\ref{Tdependence}(\textbf{e}). We stress that such subtle
spectral weight suppression is assured to be purely dependent of
temperature instead of the aging effects or noises. Besides the high
sampling points and fine statistic, at the end of the temperature
evolution study, the first spectrum is always repeated to make sure
they overlap, as testified by the two coincident 15K spectra in
Fig.\,\ref{Tdependence}(\textbf{a}).

\begin{figure}[htbp]
\includegraphics[width=8cm]{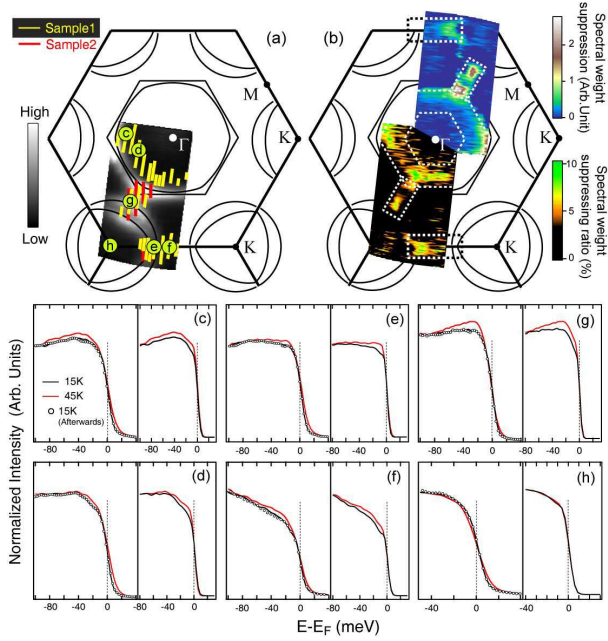}
\caption{(Color online) (\textbf{a}) The illustration of the regions
where gap opening occurs upon CDW transition in the first BZ. The
gapped regions determined by comparison of the spectra below and
above $T_{CDW}$ one by one. (\textbf{b}) The spectral weight
suppression map and its corresponding ratio map (see text for
detail) are false color plotted, where the dashed lines contour the
obviously suppressed regions and the black ones indicate the FS's.
(\textbf{c})-(\textbf{h}) Typical spectra (left) and the
corresponding thermal broadening removed ones (right) in the normal
and CDW states at momenta marked in panel (\textbf{a}).}
\label{gapmap}
\end{figure}

Following this analysis, more than 1,200 pairs of the normal and CDW
state spectra were compared one by one over more than 1/6 of the BZ.
Figs.\,\ref{gapmap}(\textbf{c-h}) give some more examples, where the
left panels show the raw spectra, and right panels show the
corresponding spectra with thermal broadening removed as practised
in Fig.2. Eventually, the momenta regions where the spectral weight
is suppressed can be determined accurately, as marked in
Fig.\,\ref{gapmap}(\textbf{a}) by bars. These regions are mainly
composed of the large domains inside the \textbf{$\Gamma$} pockets,
around the \textbf{\textbf{M}} points, and around the saddle band
points. Such distribution has been confirmed by many samples, and
the results based on two of them are shown here. Note this
distribution includes some ``barely occupied" regions where the DOS
at $E_F$ is very weak. We emphasize that the weak spectral weight
and its suppression are not due to the umklapp scattering of the CDW
order, since it is not observed in the FS or dispersion.

To further visualize the distribution of the weight suppression, the
intensity map at 15K [integrated over ($E_F$-80\,meV, $E_F$)] is
subtracted from that at 45K, and the result is shown in the upper
half of Fig.\,\ref{gapmap}(\textbf{b}). Note the thermal broadening
effects have been removed from the spectra following the standard
symmetrization method before the processing \cite{Norman}. After
normalized by the intensity map at 45K, the resulting spectral
weight suppression ratio map [lower part of
Fig.\,\ref{gapmap}(\textbf{b})] further highlights the
weight-suppressed regions, which are consistent with those
determined in Fig.\,\ref{gapmap}(\textbf{a}). Note we have surveyed
the entire BZ, and only observed suppression instead of spectral
weight gain in the measured energy range. Considering the
conservation of the total electrons, the suppressed weight should be
shifted to some higher binding energies. In this way, the electronic
energy gain for the CDW transition is naturally fulfilled.

\begin{figure}[htbp]
\includegraphics[width=8cm]{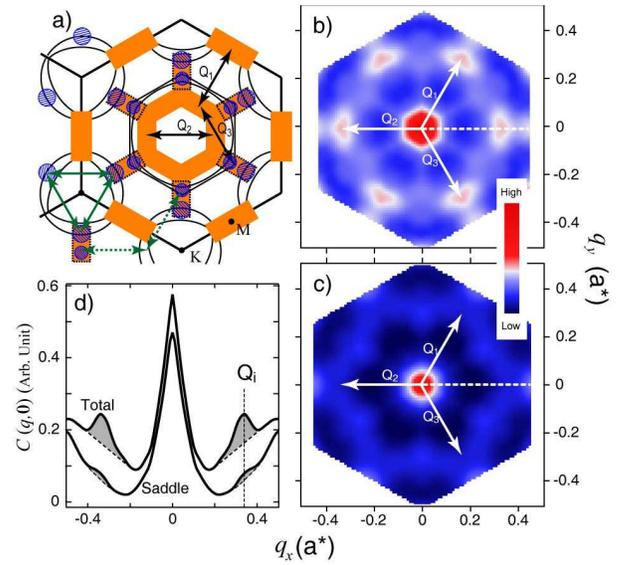}
\caption{(Color online) (\textbf{a}) The distribution of CDW related
weight suppression over the entire BZ sketched from the spectral
weight suppression map, which are marked by the orange
broken-honeycomb-shaped structure. The gapped regions could be well
connected by the CDW wavevectors, \textbf{Q$_{i}$'s}
(\emph{i}=\emph{1,2,3}), as indicated by the double-headed arrows.
The gapped regions proposed by Kiss \emph{et al.} and around the
saddle band points are marked by the hashed circles and the dotted
line surrounded rectangles. (\textbf{b})-(\textbf{c}) Two
dimensional joint DOS results on the honeycomb regions and the
saddle band point regions, respectively. (\textbf{d}) Comparison of
the corresponding joint DOS results along the dashed lines in panels
(\textbf{b}) and (\textbf{c}).} \label{Auto}
\end{figure}

The distribution of the CDW induced spectral weight suppression
makes a broken-honeycomb-shaped structure in the BZ, as shown by
Fig.\,\ref{Auto}(\textbf{a}). The suppression over the saddle band
regions (surrounded by the dotted lines) only account for 43\% of
the total one. In other words, although the DOS is higher around the
saddle band points, it only plays a secondary role in the CDW
formation due to its relatively small momentum space. Moreover, the
broken-honeycomb regions can be well connected by the CDW wavevector
\textbf{Q$_{i}$}'s [represented by the double-headed arrows in
Fig.4(a)]. In previous studies, since only a small portion of the
weight-suppressed region was discovered (hashed circles), multiple
scattering has to be invoked to barely make some of the regions
connected by the two dashed arrows in Fig.\,\ref{Auto}(\textbf{a})
\cite{Kiss}. However, this multiple scattering assumption is not
needed, once the large weight-suppressed region is recovered here.

To further demonstrate the contribution to the CDW instability of
the  broken-honeycomb-shaped region, we calculated the weighted
joint DOS,
$$C(\vec{q},\omega)\equiv\int M(\vec{k},\vec{k}+\vec{q})A(\vec{k},\omega)
A(\vec{k}+\vec{q},\omega) d\vec{k}$$, where $M=\{
\begin{array}{ll}
1 & \textrm{($\vec{k}+\vec{q}$, $\vec{k}$ $\in$ gapped region)}\\
0 & \textrm{(otherwise)}
\end{array}$. It has been shown before that $C(\vec{q},0)$ could
effectively describe the phase space for scattering of electrons
from the state at $\vec{k}$ to the state at $\vec{k}+\vec{q}$ by
certain modes with wavevector $\vec{q}$, and it is closely
associated with the electronic susceptibility
 \cite{Chatterjee,McElroy,ShenTMD}. The matrix element
$M(\vec{k},\vec{k}+\vec{q})$ is introduced assuming that only the
states that directly couple (thus suppressed) with the softening
phonon branch around \textbf{Q$_{i}$} \cite{MonctonPRB}, would
contribute to the CDW instability.  $C(\vec{q},0)$'s computed on the
honeycomb regions [Fig.\,\ref{Auto}(\textbf{b})] and on the saddle
band regions [Fig.\,\ref{Auto}(\textbf{c})] are plotted in the same
scale. Clearly, the peaks at \textbf{Q$_{i}$}'s are very prominent
in Fig.\,\ref{Auto}(\textbf{b}), whereas they are not so pronounced
in Fig.\,\ref{Auto}(\textbf{c}). This is more clearly illustrated in
Fig.\,\ref{Auto}(\textbf{d}), where the $C(q,0)$'s along the
\textbf{$\Gamma$-M} direction are compared. The saddle band regions
only contribute about 17\% of the total $C(Q_{i},0)$ computed over
the broken-honeycomb region respectively \cite{other}. Therefore,
the nesting of states with low DOS at $E_F$ indeed dominates the
charge instability at \textbf{Q$_{i}$}'s here.

% Discussion

The data here also provide some clues on the  driving force behind
the uncoventional CDW transition in NbSe$_2$. For the conventional
Peierls transition at one dimension, the CDW is driven by the
electronic energy gain, and it is reflected by a peak in the charge
susceptibility \cite{Gruner}. Although it might be more complicated
in NbSe$_{2}$, the observed peak in the weighted joint DOS does hint
that the electronic energy gain might be the dominating factor here
as well.

% ---------------conclusions -----------------

In conclusion, we have found that the CDW related spectral weight
suppression in the large broken-honeycomb region mainly away from
the FS's for NbSe$_{2}$, which naturally fulfills the nesting
condition, and resolves a long-standing puzzle of this classic CDW
material. Despite of the weak DOS in most of this region, their
primary role in the CDW formation is demonstrated. Our findings show
that collective contribution of the weakly occupied states can play
the primary role in CDW formation in some cases, and thus should not
be overlooked, which amends the conventional wisdom of CDW.

We gratefully acknowledge the helpful discussion with Profs. Q. H.
Wang, T. Xiang, J. X. Li, and H. Q. Lin. This work was supported by
NSFC, MOST (973 project No.2006CB601002 and No.2006CB921300), and
STCSM of China.


\begin{references}
\bibitem{other} Alternatively, if estimated based on the height, the ratio is about 28\%.
\end{references}
\end{document}